\documentclass[12pt]{article}
\usepackage[utf8]{inputenc}
\usepackage{amsmath}
\usepackage{graphicx}
\usepackage{caption}
\usepackage{hyperref}
\usepackage{subcaption}
\usepackage{algorithm}
\usepackage{algpseudocode}
\usepackage{listings}
\usepackage{xcolor}
\usepackage[normalem]{ulem}

\title{Cooperative Behavior in Pre-State Societies: An Agent-Based Approach of the Aksum Civilization}

\author{Riccardo Vasellini$^{1,2}$, Gilda Ferrandino$^{3}$, Luisa Sernicola$^{3}$, \\ Daniele Vilone$^{1,4}$
 and  Chiara Mocenni$^{2}$ \\
\ \ \ \\
$^1$LABSS-ISTC, National Research Council, Rome, Italy; \\
$^2$DIISM, University of Siena, Siena, Italy; \\
$^3$Universit\`a degli Studi di Napoli “L’Orientale”; \\ 
$^4$GISC-Dep.to de Matem\'aticas, Universidad Carlos III de Madrid, Spain 
}

\date{}

\begin{document}

\maketitle

\begin{abstract}
  This study intends to test the hypothesis that, contrary to traditional interpretation, the social structure of the polity of Aksum -- especially in its early stages -- was not characterized by a vertical hierarchy with highly centralized administrative power, and that the leaders mentioned in the few available inscriptions were predominantly ritual leaders with religious rather than coercive political authority. This hypothesis, suggested by the available archaeological evidence, is grounded in Charles Stanish’s model, which posits that pre-state societies could achieve cooperative behavior without the presence of coercive authority. Using agent-based modeling applied to data inspired by the Aksum civilization, we examine the dynamics of cooperation in the presence and absence of a Public Goods Game. Results show that while cooperative behavior can emerge in the short term without coercive power, it may not be sustainable over the long term, suggesting a need for centralized authority to foster stable, complex societies. These findings provide insights into the evolutionary pathways that lead to state formation and complex social structures.
\end{abstract}

\section{Introduction}

The evolution of complex societies has often been linked to the emergence of centralized, coercive authority capable of maintaining order and cooperation among large populations\cite{Carneiro}. Charles Stanish proposed that pre-state societies could exhibit cooperative behaviors without the presence of coercive power\cite{Sta17}, but that sustained cooperation likely requires some form of centralized authority. In this study, we explore this hypothesis by simulating a pre-state society inspired by data from the Aksum civilization using an agent-based model. The model includes a Public Goods Game (PGG) to simulate cooperative behavior, aiming at understanding the conditions under which cooperation can emerge and be sustained without coercive mechanisms.

\ 

Situated in the mountainous regions of the northern Ethiopian plateau, from 
the second half of the 1st millennium BCE the area of Aksum progressively 
emerged among the polities of the northern Horn of Africa, extending 
throughout the 1st millennium CE its influence over an increasingly vast 
territory, which reached across the Red Sea to include part of present-day 
Yemen\cite{phillipson2012}. Throughout its history, the so-called ‘Kingdom of Aksum’ was involved in 
long-distance trade and contacts with the Mediterranean, the Nile Valley, the 
Arab side of the Red Sea and the western Indian Ocean\cite{manzo2005, Munro82, Munro96}. It reached 
sophisticated levels of architectural expertise, one thinks, for example, of the 
great funerary stelae, some of which are among the largest monoliths ever 
erected in the world \cite{phillipson1977excavation}. It minted, a unique case in ancient Africa, its own 
currency with denominations in bronze, silver and gold (although, as the 
archaeological evidence suggests, the latter two were mainly used abroad and 
part of the local population still practiced barter as the main exchange system)\cite{munrohay1991}. 
Last but not least, it played a significant role in the geopolitical scenario of the
Near East in the Late Antique period\cite{bowersock2013}. 
Despite this, there is very little written documentation referable to this period 
in Aksum's history, most consisting of inscriptions commemorating military raids 
or other achievements by chiefs\cite{fattovich2010development}. There are very few references to a 
bureaucratic/administrative apparatus, essential for the alleged centralised 
management of such a vast and diversified territory. For the earlier phases, 
written documentation is even poorer and inscriptions remain essentially 
celebratory and ritual in content\cite{bernand1991recueil}. Although inscriptions mention epithets 
translated as ‘king’, their number and content as well as the available 
archaeological evidence do not support, especially for the earliest phases, the 
traditional interpretation of these social entities as kingdoms or states, at least 
in the common understanding of the terms. 
Based on this, our paper discusses the preliminary results of an ABM-based 
study on the nature of the earliest forms of complex societies in this area and 
its possible development trajectories over time. It relies on firm archaeological 
and ethnographic data and explores concepts of clanship, cooperation, 
ritualized economy and leadership derived from cultural anthropology and 
allied disciplines~\cite{mcintosh1999}.

\section{Methodology}

\textit{This paper proposes the development of an ABM aimed to describe the cooperative behavior of the Aksum population.
This section describes the studies done by archaeologists and  the mathematical model over which is made. % The model is built with strict adherence to the data that will described in the following subsection.% 
} 
%\subsection{Archaeological Studies and Data Acquisition}
%\textbf{SPIEGARE COME SONO STATI EFFETTUATI GLI STUDI ARCHEOLOGICI E  COME SONO STATI ACQUISITI I DATI.}

\subsection{Archaeological Studies and Data Acquisition}
Due to its historical relevance, Aksum has been undoubtedly one of the most thoroughly archaeologically investigated sites of the northern Horn of Africa. Research in the area include both archaeological excavations and survey, which increased knowledge on the dynamics that underwent ancient Aksum’s emergence and subsequent development. Alongside data available in the scientific literature, two intensive archaeological survey projects conducted between 2005 and 2006 in the framework of the Italian Archaeological Expedition at Aksum of the University of Naples L'Orientale and the World Bank ‘Ethiopian Cultural Heritage Project’ concurred to provide a GIS-based archaeological map of the whole area of Aksum. The map includes an overall amount of 698 sites (settlements, monumental buildings, cemeteries, landscape infrastructures) distributed across an area of about 105 sq km~\cite{Sernicola17}. Geological, geoarchaeological and environmental studies enabled the generation of thematic maps on soil productivity, water resources distribution and slope gradient. This formed the basis for the reconstruction of ancient occupation dynamics, demographic trend and land-exploitation strategies in the area. GIS analysis showed that since the first millennium BCE, settlements were located according to three major environmental factors: a) close proximity to water resources (no more than 250 m from rivers, streams or water reservoirs), b) close proximity to more productive soil, and c) slope gradient~\cite{Sernicola17}.
\subsection{Agent-Based Model}

An agent-based model was developed to simulate the cooperative behavior of agents %(turtles) IO EVITEREI DI CHIAMARE GLI UGENTI "TURTLES" E LE RISORSE "SUGAR" NEL PAPER
inspired by the Aksum civilization. The model includes agents with properties such as resource levels, %(sugar),
metabolic needs, vision, and a probability of contributing to the public good. The PGG was implemented to simulate cooperative interactions, where agents could contribute a portion of their resources to a common pool, which in turn benefited the whole community if the collective donation met a defined threshold. Fig 1 reports the initialization of the model where the orange areas represent the proto Aksumite sites, the green the fertility, and the cyan the river in strict coherence with the knowledge emerging from archaeological studies. 
\begin{figure}[!ht]
    \centering
    \includegraphics[width=0.5\textwidth]{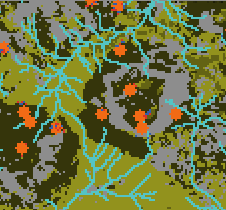}
    \caption{Initial state of the NetLogo simulation. The orange circles represent areas where the Public Goods Game (PGG) can be performed, centered at proto-Aksumite sites. Grey patches have zero fertility, while the green gradient represents varying fertility levels across the map. The cyan curves are rivers.}
    \label{fig:netlogo-beg}
\end{figure}

Our agent-based model integrates two complementary sets of features. First, we ground the spatial and demographic aspects of the simulation in high-resolution archaeological and GIS data from the Aksum region. This empirical foundation ensures that basic environmental constraints and initial settlement patterns closely mirror the real landscape of northern Ethiopia around the first millennium BCE. Second, we overlay a suite of theoretical mechanisms—drawn from evolutionary game theory and ecological modeling-that allow us to investigate how cooperation might emerge, persist, or fall apart in a pre-state society.

In practice, this dual approach means:

\paragraph{Empirical  Components}
These elements are directly informed by archaeological evidence and spatial analyses:
\begin{itemize}
  \item \textit{GIS-derived fertility and hydrology layers:} Soil productivity (“fertility” values) and river networks imported from shapefiles, defining patch regeneration rates and movement barriers~\cite{Sernicola17}.
  \item \textit{Site coordinates:} Locations of 698 pre- and proto-Aksumite settlements determine where initial agents are placed and where Public Goods Games can occur.
  \item \textit{Settlement heuristics:} Empirical findings that ancient villages clustered within 250\,m of water sources and on productive soils shape the placement and mobility of agents~\cite{Sernicola17}.
\end{itemize}

\paragraph{Theoretical  Components}
These rules and parameters are designed to probe cooperation dynamics beyond what the archaeological record alone can reveal:
\begin{itemize}
  \item \textit{Public Goods Game framework:} Donation rates, dynamic thresholds, and benefit multipliers allow us to study how local cooperation might boost resource regeneration.
  \item \textit{Movement and vision algorithms:} Sugarscape-inspired heuristics position agents to seek high-fertility patches, testing how competition for resources interacts with social behavior.
  \item \textit{Reproduction and mutation:} Resource-based birth rules and stochastic inheritance of donation probability explore demographic feedback on cooperation.
  \item \textit{Resource degradation/regrowth cycles:} Simple decay and renewal functions capture environmental feedback loops that can destabilize or reinforce cooperative clusters.
\end{itemize}

Table~\ref{tab:parameters} then lists every parameter and indicates whether it is data-driven or theoretical, making the model’s hybrid nature explicit.

%\newline

The model was designed with two scenarios:
\begin{itemize}
    \item \textbf{Scenario 1:} Public Goods Game (PGG) enabled.
    \item \textbf{Scenario 2:} No Public Goods Game.
\end{itemize}

\subsection{Public Goods Game}

%In PGG rounds, cooperators contribute a given quantity $c>0$, whereas defectors contribute nothing. Then, the total contribution collected by the group is multiplied by a factor $B$ and equally shared among all group members, regardless of individual contributions.

In classical Public Goods Games (PGGs), each agent in a group chooses whether to contribute a cost $c > 0$ to a common pool. The total contribution is multiplied by a synergy factor $r$ and then redistributed equally among all group members, regardless of contribution\cite{rot95}. The payoff for each agent $i$ in a group of size $G$ with $N_c$ cooperators is thus:
\[
\Pi_i = \begin{cases}
\frac{r \cdot N_c \cdot c}{G} - c & \text{if } i \text{ is a cooperator} \\
\frac{r \cdot N_c \cdot c}{G} & \text{if } i \text{ is a defector}
\end{cases}
\]

In spatial extensions of the PGG, agents are placed on a grid and interact within a local neighborhood, typically defined by a Moore or von Neumann structure \cite{szabo2007evolutionary}. Each agent plays multiple overlapping games with its neighbors, and the results depend on the spatial arrangement of cooperators and defectors . These models assume a fixed neighborhood and no agent movement.

Our model departs from this structure in two ways. First, agents move freely across space and can enter or leave cooperative zones. Second, rather than receiving individual payoffs, cooperation affects the environment itself by increasing patch fertility. In this sense, the "payoff" is externalized into the landscape.

\subsubsection*{The Mechanism of PGG in the Aksum Model}

\begin{enumerate}
    \item \textbf{Identification of PGG Centers:}
    The PGG is initiated at specific geospatial locations identified as proto-Aksumite site patches. These patches represent areas of early human settlement or resource concentration, loaded from GIS data.

    \item \textbf{Participation Radius:}
    Each PGG center defines a radius within which agents (turtles) are eligible to participate. This radius represents the spatial influence of the PGG center and reflects the localized nature of resource sharing and cooperation in early societies.

    \item \textbf{Threshold Calculation:}
    The threshold for a successful PGG is dynamically determined based on:
        \begin{enumerate}
            \item A population-based threshold proportional to the number of agents within the radius.
            \item A resource-based threshold calculated as a fraction of the total resources (sugar) of the participating agents.
        \end{enumerate}
        Mathematically, the success threshold $T$ is:
        \[
        T = \alpha \cdot N + \beta \cdot R
        \]
        where:
        \begin{itemize}
            \item $N$ is the number of agents in the participation radius
            \item $R$ is the total sugar of these agents
            \item $\alpha$ and $\beta$ are constant scaling parameters (e.g., 0.05 each)
        \end{itemize}
        
Note that in the code (and pseudocode) there is a third component for the threshold, this is an element used to avoid bugs in the execution of the code that avoids PGG with threshold 0 to happen. 
    \item \textbf{Agent Behavior:}
    Agents within the participation radius evaluate their willingness to contribute based on their donation probability. In our model, donation is a binary trait: agents are either full cooperators or full defectors. If cooperative, an agent donates a fixed fraction (e.g., 10\%) of its resources. Defectors donate nothing.

    \item \textbf{Outcome of the PGG:}
    If the total donation $D$ meets or exceeds the threshold $T$:
        \begin{itemize}
            \item The PGG is deemed successful.
            \item Each patch within the participation radius increases its maximum fertility by a fixed amount $\Delta_f$ (e.g., 1.5), capped at a global ceiling.
        \end{itemize}
    If the threshold is not met, no benefit is conferred.
\end{enumerate}

\subsubsection*{Environmental Update Equation}

Regardless of success or failure, fertility undergoes degradation over time (see section 2.5). At each tick, patch $i$ updates its fertility according to:
\[
F_i(t+1) = \min\left[
\max\left(F_i(t) - \gamma, F_{\text{base}}\right) + \delta_i(t),\,
F_{\text{cap}}
\right]
\]

where:
\begin{itemize}
    \item $F_t$ is the fertility at time t
    \item $\gamma$ is the degradation decrement (e.g., 0.7)
    \item $F_{\text{base}}$ is the baseline fertility (patch-specific)
    \item $F_{\text{cap}}$ is the maximum fertility ceiling (e.g., 100)
    \item $\delta_i(t) = \Delta_f$ if patch $i$ is in a successful PGG radius at time $t$, otherwise 0
\end{itemize}

This ensures that cooperation can temporarily boost fertility, but such gains decay in the absence of sustained collective action.
For more details we provided the pseudo code of the mechanisms of the PGG in the Appendix (Algorithm 1).

\subsubsection*{Key Differences from Standard PGG Frameworks}

The PGG mechanism in the Aksum model differs from traditional implementations in the following ways:

\begin{enumerate}
    \item \textbf{Spatial Context:}
    Traditional PGG models typically assume a non-spatial environment where participants interact globally or within fixed networks. In contrast, the Aksum model incorporates geospatial data to determine participation areas and thresholds, anchoring the game in a physical and historical context. Agents can enter or leave PGG zones, allowing for fluid, open-ended participation.

    \item \textbf{Dynamic Thresholds:}
    Unlike standard models with static thresholds, the Aksum model thresholds adapt to the local population and resource availability, making the game more representative of real-world conditions.

    \item \textbf{Localized Benefits:}
    The benefits of cooperation (fertility increases) are confined to the spatial area surrounding the PGG center, emphasizing the local impact of collective actions. This approach mirrors the dynamics of resource sharing in pre-state societies.

    \item \textbf{Integration with Environmental Feedback:}
    The Aksum model links the success of the PGG to environmental feedback by adjusting the fertility of patches. This coupling highlights the interplay between social cooperation and ecological sustainability, a feature absent in many classical PGG models.
\end{enumerate}

\subsection{Reproduction, Movement, and Sugarscape Foundations}

Our model builds on the foundational Sugarscape framework~\cite{epstein1996growing} (see Appendix, Algorithms 2-5) , in which agents collect resources (called "sugar") from a landscape, consume a fixed amount per time step (their metabolism), and die if their resources run out or if they reach a maximum age. Agents move to nearby patches with the most sugar, and patches regenerate sugar each tick up to a predefined fertility maximum. This setup allows basic ecological dynamics like scarcity, competition, and reproduction to emerge from simple rules.

We retain these core elements—including metabolism, vision-based movement, and resource-based reproduction—as the ecological backbone of the Aksum model. On top of this, we add new mechanisms (such as the Public Goods Game) to simulate collective action and cooperation in pre-state societies.

The Aksum model incorporates reproduction and movement mechanisms to reflect dynamic population growth and resource-seeking behaviors, adding complexity to the simulation.

\subsubsection{Reproduction Mechanism}
\begin{itemize}
\item \textbf{Resource-Based Reproduction:} Agents can reproduce if they accumulate enough resources (sugar) to meet a predefined threshold, typically set as twice their metabolic requirements. This ensures only resource-rich agents contribute to population growth.
\item \textbf{Offspring Inheritance:} Offspring inherit only the donation probability, with a small probability of mutation to introduce variability. Every other trait, such as vision range and metabolism, is fixed in order to isolate evolutionary pressure to social behavior.
\item \textbf{Resource Sharing:} Parents share their resources equally with offspring, temporarily reducing their own resource pool to simulate reproduction costs.
\item \textbf{Reproduction Limit:} Agents are capped at a maximum number of reproductions to maintain a balance between population growth and resource availability.
\end{itemize}

\subsubsection{Movement Mechanism}
\textbf{Vision-Based Movement:} Agents scan their environment within a defined vision range for patches with higher fertility. Then they evaluate destinations based on resource availability and the number of other agents present, reducing competition and preventing overloading of fertile patches. Two agents cannot occupy the same patch at the same time.

\subsection{Regrowth and Degradation of Resources in Patches}

Resource dynamics in patches are modeled through processes of regrowth and degradation, which simulate environmental feedback and sustainability. The regrowth mechanism follows Sugarscape’s standard rule, where each patch regenerates its resources up to a fixed maximum per tick. However, we extend this by introducing a degradation mechanism: when cooperation increases fertility beyond the baseline, this improvement decays over time in the absence of further cooperative action. This models ecological fragility, a feature not present in the original Sugarscape. The increment of fertility as a consequence of successful PGGs will always be higher than the degradation applied. This is to say that if there's cooperation, fertility grows (despite degradation), while if there's no cooperation, the fertility increments accumulated in the previous tick will revert to the original baseline of the patch after a while if no PGGs are successful.

\subsection{Model Assumptions and Parameters}

The parameters used in the model are based on a combination of archaeological evidence and theoretical assumptions.  
Where possible, values were informed by GIS data and archaeological surveys (e.g., soil fertility, settlement locations).  
When no direct data were available (such as the donation rate or reproduction thresholds), parameters were chosen to be qualitatively plausible and were further tested via sensitivity analyses.  Table~\ref{tab:parameters} summarizes the parameters, their interpretation, and their value ranges.

\begin{table}[H]
\vspace*{-3cm}
  \centering
  \small
  \setlength{\tabcolsep}{6pt}
  \renewcommand{\arraystretch}{1.2}
  \caption{Model parameters: values, units, and interpretations.}
  \label{tab:parameters}
  \begin{tabular}{|p{4cm}|p{3cm}|p{8cm}|}
    \hline
    \textbf{Parameter} & \textbf{Value (Unit)} & \textbf{Description and Rationale} \\
    \hline
    Donation rate & 0.10 (fraction) & Assumed plausible fraction of resources donated by cooperating agents at each PGG; tested sensitivity range 0.10–0.50 \\
    \hline
    PGG population factor & 0.05 per agent (sugar) & Hypothetical scaling of PGG threshold with local group size. Calibrated, along with the resource factor, in order to need, on average, more than half of the agents in a PGG area to be cooperators for the PGG to be successful with a donation rate of 0.1. \\
    \hline
    PGG resource factor & 0.05 (fraction) & Hypothetical scaling of PGG threshold with total agent resources in the PGG radius \\
    \hline
    PGG participation radius & 3 (patches) & Assumed plausible based on settlement proximity in GIS survey data \\
    \hline
    Vision range & 5 (patches) & Inspired by plausible human mobility limits at the modeled scale \\
    \hline
    Metabolism & 2 (sugar/tick) & Assumed resource consumption rate per time step; tested sensitivity range 1-4 \\
    \hline
    Patch fertility  & 0-4 (sugar) &  
    It is the maximum sugar a patch can ever hold, can be increased by successful PGG. Initial values are seeded from the 30m‑resolution GIS soil‑productivity raster; Capped at 100.     Implemented in the code as \texttt{max\_pSugar}.  \\
    \hline
    Degradation fertility decrement & 0.7 (fertility)  & Assumed decrement of fertility applied each turn to patches with higher fertility than their initial value.\\
    \hline
    PGG fertility increment & 1.5 (fertility) & Fixed bonus added to each patch’s maximum fertility whenever a Public Goods Game succeeds, ensuring cooperation yields a net gain that outweighs one‑tick degradation.\\
    \hline
    Reproduction threshold & $2 \times$ metabolism (sugar) & Model rule: reproduction allowed after storing twice metabolic cost \\
    \hline
    Max age & Uniform[40,60] (ticks) & Agent lifespan variability reflecting demographic uncertainty \\
    \hline
    Reproduction cap & 2 (offspring) & Max number of births per agent to moderate population explosion \\
    \hline
    Mutation rate & 0.05 (probability) & Probability of changing cooperation strategy across generations \\
    \hline
    Runs per scenario & 100 (runs) & Statistical robustness for scenario comparison \\
    \hline
  \end{tabular}
\end{table}

\subsection{Statistical Analysis}

To assess the statistical significance of differences in population dynamics over time between the two scenarios, we conducted a time-series analysis.  
At each simulation tick, a \textbf{Welch's t-test} was applied between the PGG and no-PGG conditions \cite{welch1947generalization}.  
Because multiple comparisons were performed across all time steps, we applied a \textbf{Bonferroni correction} to control for the familywise error rate \cite{Miller}.  
The adjusted significance threshold was calculated as
\[
\alpha_{\text{adjusted}} = \frac{0.05}{N}
\]
where $N$ is the number of time steps.
\section{Results}
In this section we report the results of the ABM regarding spatio-temporal dynamics and population dynamics. We talk about their significance and their adherence to the archaeological findings. 

\subsection{Spatiotemporal Dynamics}
\begin{figure}[h!]
    \centering
    \includegraphics[width=0.5\textwidth]{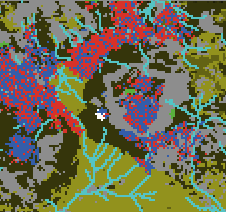}
    \caption{Clustered state of the simulation. We observe clusters of cooperating agents (blue) forming around a proto-Aksumite site in the bottom-left and another near the center of the map.}
    \label{fig:netlogo-cluster}
\end{figure}

As the simulation progresses, clusters of cooperators emerge  around the proto-Aksumite sites. In Figure~\ref{fig:netlogo-cluster} we observe two notable clusters of cooperating agents (blue): one in the bottom-left and another near the center of the map. These clusters suggest that cooperation is initially possible in localized areas where the PGG is actively performed.

\begin{figure}[h!]
    \centering
    \includegraphics[width=0.5\textwidth]{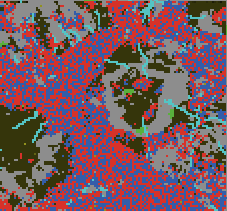}
    \caption{End state of the simulation. Clusters of cooperators have disappeared, and defectors (red agents) dominate the map. However, cooperators do not vanish entirely due to specific genetic and spatial dynamics.}
    \label{fig:netlogo-end}
\end{figure}

However, by the end of the simulation (Figure~\ref{fig:netlogo-end}), clusters of cooperators are no longer present. Defectors (red agents) make it impossible for the PGG to be successfully performed. Despite this, cooperators do not disappear entirely. This is because genetic pressure favoring cooperation is applied only within the orange circles (PGG areas, Figure~\ref{fig:netlogo-beg}). Outside these areas, cooperators behave identically to defectors, effectively acting as simple Sugarscape agents.

This outcome highlights the dynamics of cooperation and defection in the system. While cooperation can emerge under favorable conditions in specific areas, defectors tend to invade and destabilize these clusters over time. 

\subsection{Population Dynamics}

\begin{figure}[!ht]
    \centering
    \includegraphics[width=0.7\textwidth]{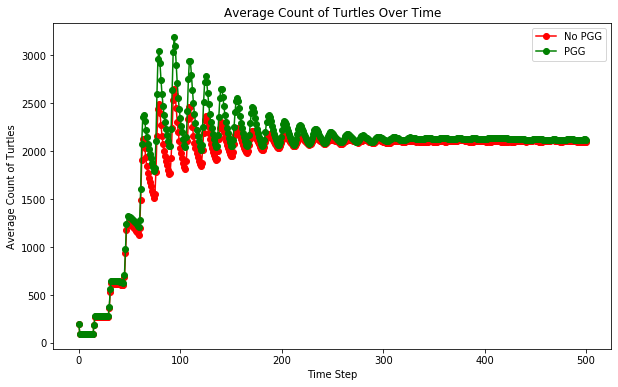}
    \caption{Average agent population over time in scenarios with and without the Public Goods Game (PGG). Results averaged over 100 independent realizations.}
    \label{fig:population-dynamics}
\end{figure}

Figure \ref{fig:population-dynamics} shows the average agent population over time in two conditions: one with the public-good game enabled (PGG, green), and one without it (No PGG, red). Both simulations display the same broad dynamic pattern: an initial exponential growth, followed by oscillations, and eventual stabilization.

These oscillations—present in both runs—are characteristic of Sugarscape-type models, where agents consume local resources, reproduce, then experience crashes as resource availability temporarily falls below metabolic demand. Over time, the system self-regulates, and population stabilizes once a balance between regeneration and consumption is reached.

The key difference lies in amplitude and peak. In the PGG scenario, early cooperation boosts local fertility through successful public-good contributions, allowing the population to grow faster and peak higher (around 3,200 agents) than in the No PGG case (which peaks at ~2,400). However, this cooperative advantage is not sustained: as the population increases, donation becomes costlier, fewer agents contribute, and PGG success declines. As a result, the fertility boost collapses, and the PGG population converges back to the same equilibrium level (~2,100) as the non-cooperative baseline.

This supports our core argument: cooperation can emerge spontaneously and temporarily raise collective outcomes, but in the absence of additional mechanisms (e.g., punishment, redistribution, coercive leadership), it tends to break down. This mirrors hypotheses in the archaeological and institutional literature that early cooperation likely needed reinforcement to become stable and enduring~\cite{Ostrom, Turchin}.

\subsection{Adherence of the Results with Archaeological Data}
The demographic trend emerged from the simulation in the PGG scenario reflects, from a qualitative point of view, what we observe in the archaeological record during the early phases of development of Aksum (i. e. throughout the 1st millennium BCE). In this period, for which we have no clear evidence of the presence of leaders exercising coercive power and a political/administrative authority, the settlement system suggests a significant demographic increase to a peak in the very first centuries of the 1st millennium CE~\cite{Sernicola17}. From this time onwards, demographic trends in the Aksum area reached a phase of general equilibrium, with small fluctuations and changes in the spatial distribution of settlements rather than in their number. This is the period in which we observe in the archaeological record and in the few available written texts the appearance of traits attributable to the leaders' acquisition of a progressively more political/administrative role (e.g. the beginning of a system of coinage, evidence of donations of land by the king to religious institutions, divine descent of the king, etc.). In this sense, the archaeological data confirms what emerged from the model: the formation of coercive authority helps to keep the system stable by managing cooperation within a larger population. This avoids the population decline observed in the simulation once the threshold of the number of agents willing to cooperate without the presence of a political/administrative authority is reached. 
The difference between the case-study of Aksum and the simulation is, rather, a quantitative one and is evident above all in the rapidity of the demographic growth observed in the model compared to the archaeological evidence. It will therefore be interesting to reason about the factors that may in reality have contributed to slower population growth. Although the variables and parameters that could have influenced this aspect may be various, an ecological crisis can be ruled out based on environmental data~\cite{Southall}. Rather, a higher infant mortality rate or a lower number of individuals per household than assumed based on the ethnographic record may have played a significant role in this respect, as well as a greater mobility of individuals, or, more likely, a major social and cultural changes as attested in this region around the mid-1st millennium BCE. These aspects will be further investigated through targeted archaeological activities in the field and future developments of the ABM simulation, together with the introduction, already in the earliest phases, of ritual leaders/mediators as suggested by the epigraphic record.

\subsection{Evolution of Cooperative Behaviour Over Time}

\begin{figure}[ht]
  \centering
  \begin{subfigure}{.48\linewidth}
      \includegraphics[width=\linewidth]{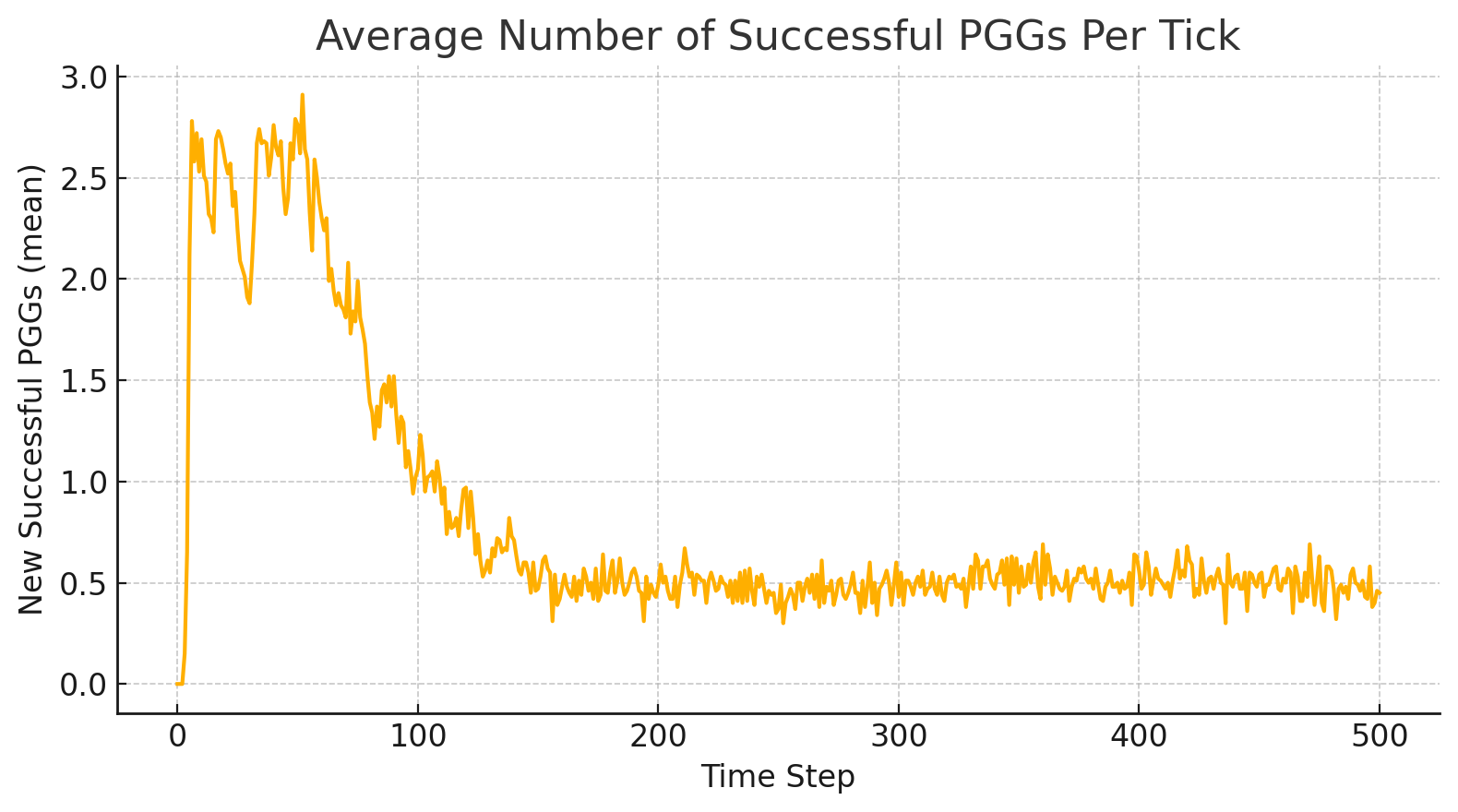}
      \caption{}
  \end{subfigure}\hfill
  \begin{subfigure}{.48\linewidth}
      \includegraphics[width=\linewidth]{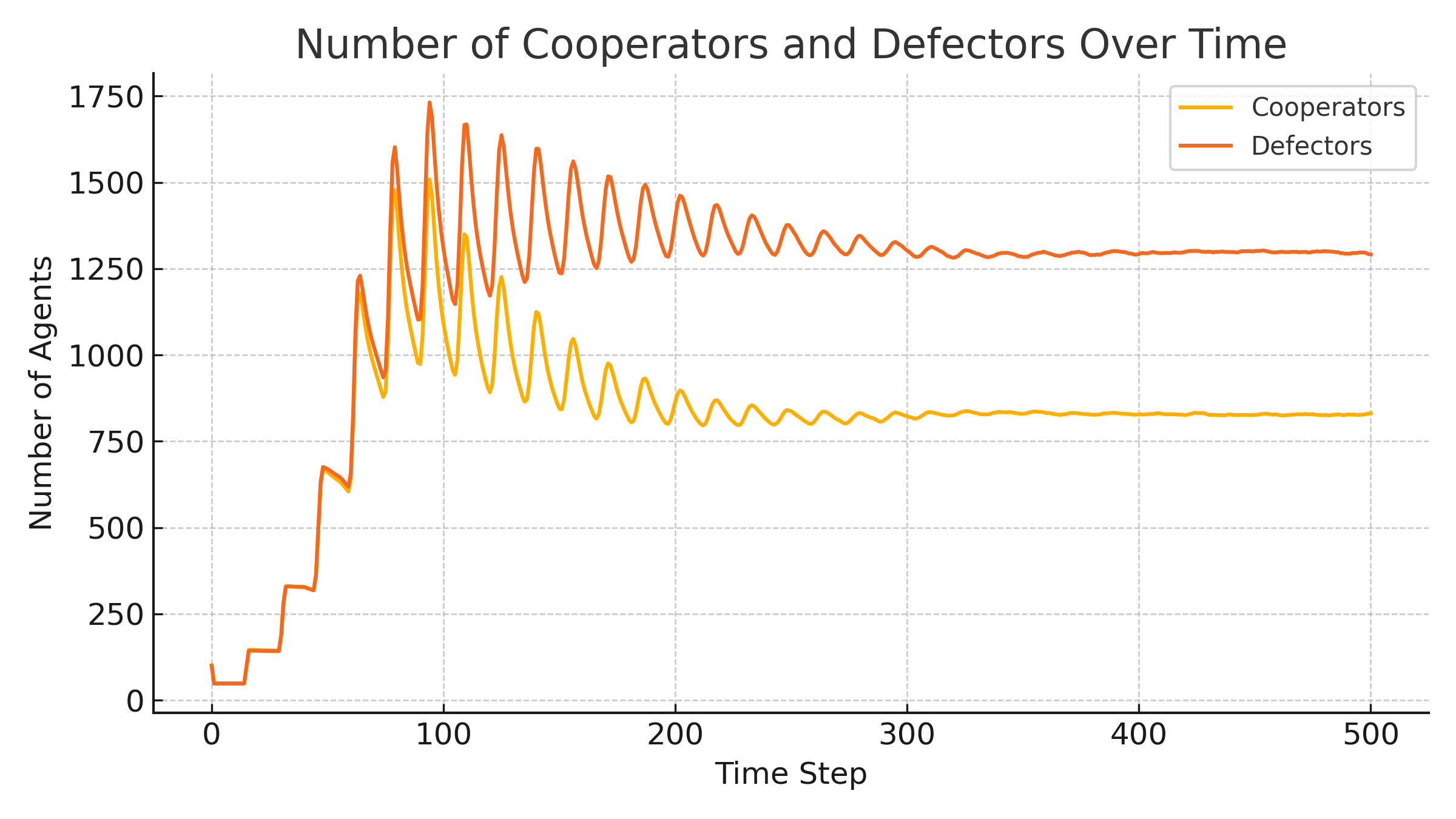}
      \caption{}
  \end{subfigure}
  \caption{Temporal dynamics of cooperation. Averages over 100 runs.  (a) Mean number of successful PGG at each tick; (b) Number of cooperators and defectors over time.}
  \label{fig:coop-dynamics}
\end{figure}

We track cooperation with two complementary indicators:

\begin{enumerate}
  \item \textbf{Incremental PGG successes} – the number of \emph{newly}
        successful Public‑Good Games each tick (Figure~\ref{fig:coop-dynamics}a).
  \item \textbf{Strategy head‑counts} – the absolute number of cooperating
        (yellow) and defecting (orange) agents each tick
        (Figure~\ref{fig:coop-dynamics}b).
\end{enumerate}

Together they reveal a three‑phase pattern:

\begin{itemize}
  \item \textbf{Ramp‑up phase (ticks 0–100).}  
        Successful PGGs rise steeply, peaking around tick 80.
        Co‑operator numbers grow in step with total population.

  \item \textbf{Decline phase (ticks 100–300).}  
        As resource competition intensifies, defectors infiltrate cooperative
        zones.  Incremental PGG successes fall sharply and co‑operator head‑count
        drops to about 40\% of the population.

  \item \textbf{Late equilibrium (ticks~300–500).}  
        A stable but lower level of cooperation persists
        ($\sim$850 cooperators vs.\ $\,\sim$1300 defectors),
        mirroring the damped population oscillations.
\end{itemize}

The figures show that cooperation \emph{persists} as a substantial minority
rather than disappearing; however, the reduced co‑operator pool cannot sustain
the early fertility boost, so total population and PGG activity settle at a much
lower plateau.  This supports our claim that, without enforcement or
redistribution, spontaneous cooperation in a growing pre‑state society is
inherently fragile.

\subsection{Statistical Significance of  Population Dynamics}

\begin{figure}[!ht]
    \centering
    \includegraphics[width=0.7\textwidth]{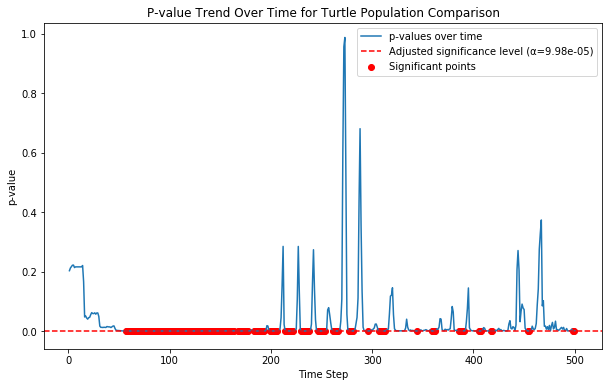}
    \caption{P-value trend over time comparing agent populations in scenarios with and without the Public Goods Game (PGG). Points below the adjusted significance threshold indicate statistically significant differences.}
    \label{fig:p-values}
\end{figure}

Figure~\ref{fig:p-values} displays the p-value trend over time.  
The horizontal dashed line represents the adjusted significance threshold.  
Time points where the p-value falls below this threshold (highlighted by red dots) indicate intervals where the agent populations under the PGG and no-PGG conditions are statistically significantly different.

Early in the simulation, the PGG condition sustains significantly higher agent populations compared to the no-PGG condition, supporting the hypothesis that Public Goods Games promote short-term cooperation and demographic success.  
However, as time progresses and cooperative behavior collapses, the difference between the scenarios diminishes, and the p-values rise above the significance threshold.

This result illustrates that, although cooperation can initially sustain larger populations without coercive institutions, its effect is transient and fragile over long-term dynamics.

\subsection{Sensitivity analysis}

\begin{figure}[ht]
  \centering
  \begin{subfigure}{.5\linewidth}
    \centering
    \includegraphics[width=\linewidth]{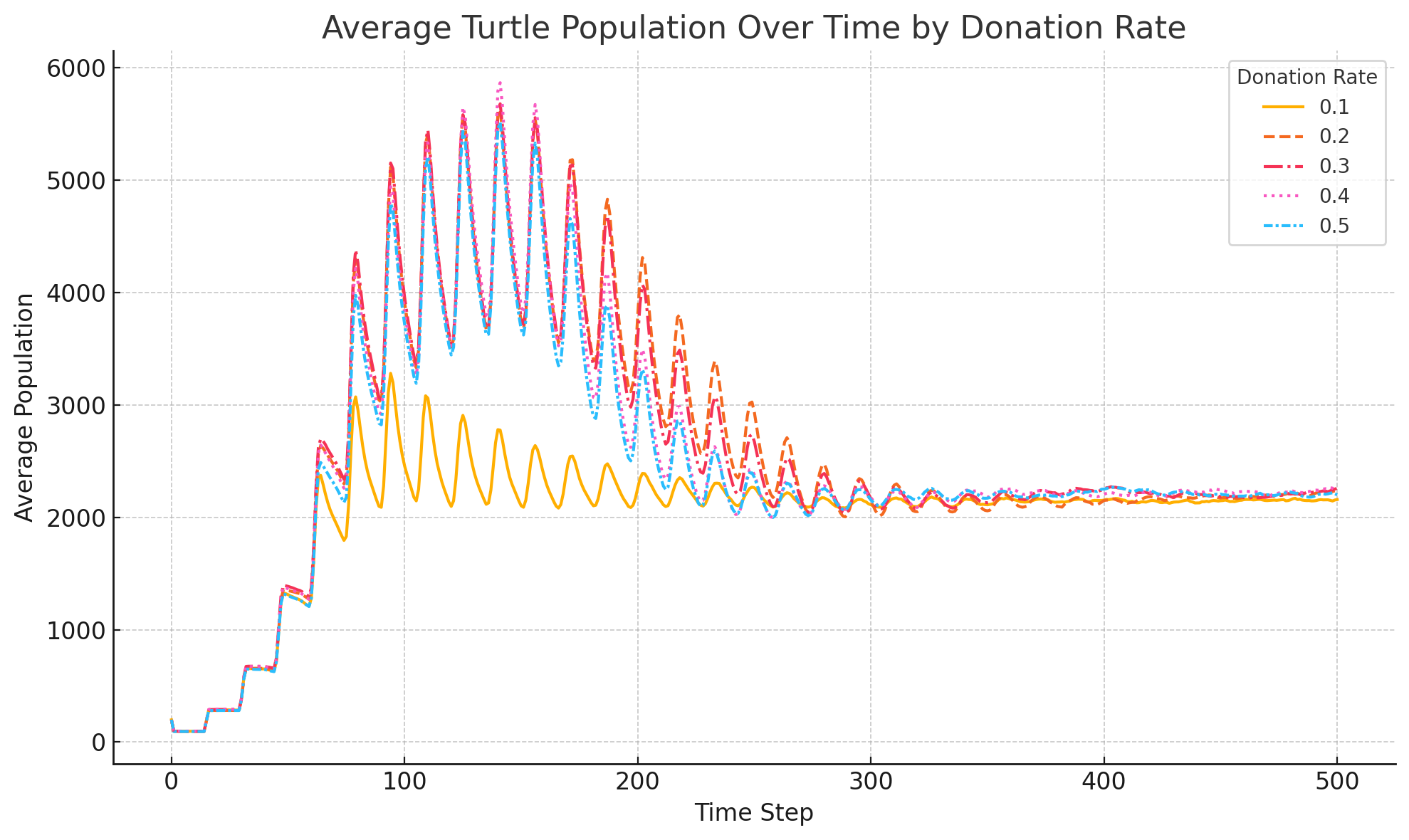}
    \caption{Donation‑rate}
  \end{subfigure}\hfill
  \begin{subfigure}{.5\linewidth}
    \centering
    \includegraphics[width=\linewidth]{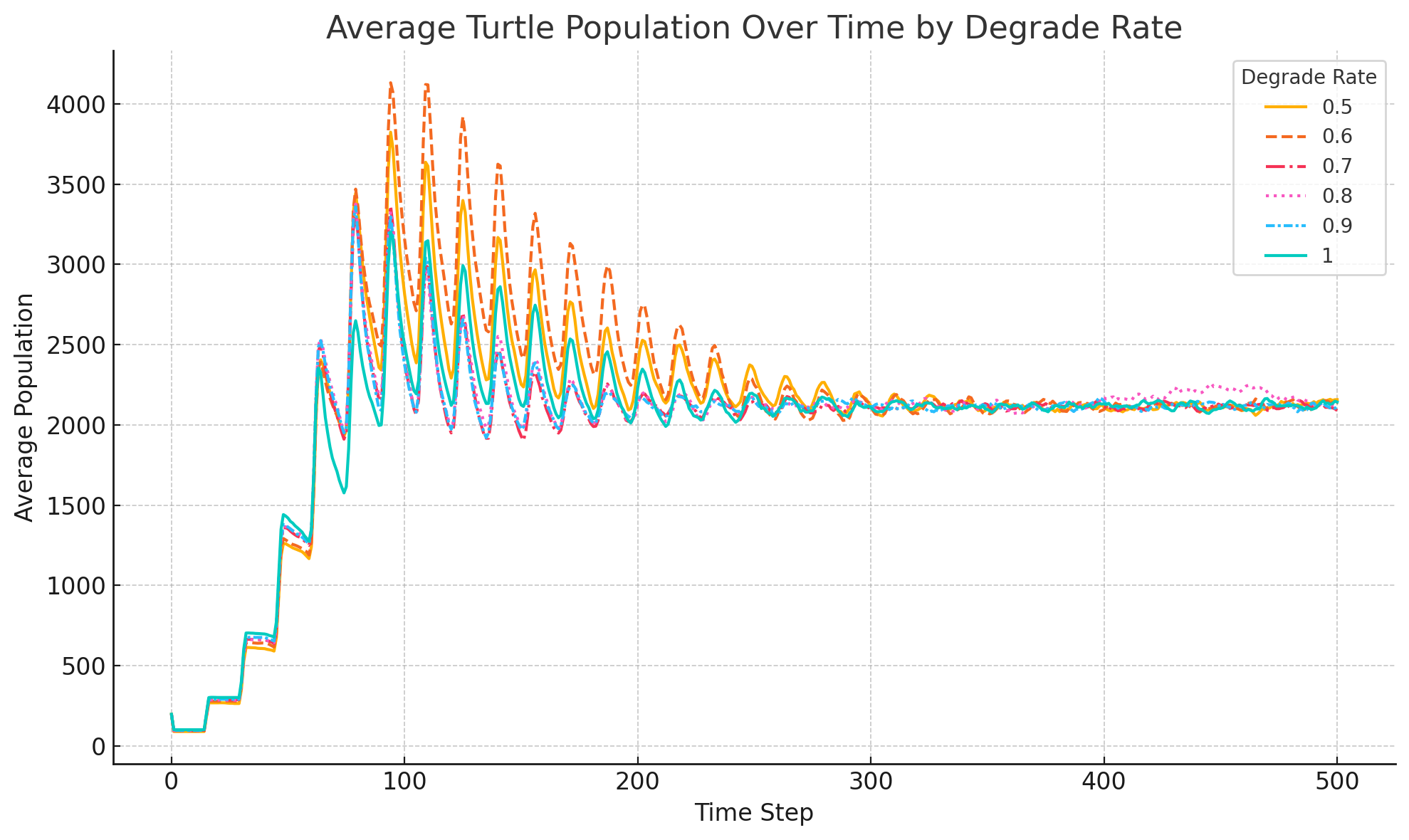}
    \caption{Degrade‑rate }
  \end{subfigure}\hfill
  \begin{subfigure}{.5\linewidth}
    \centering
    \includegraphics[width=\linewidth]{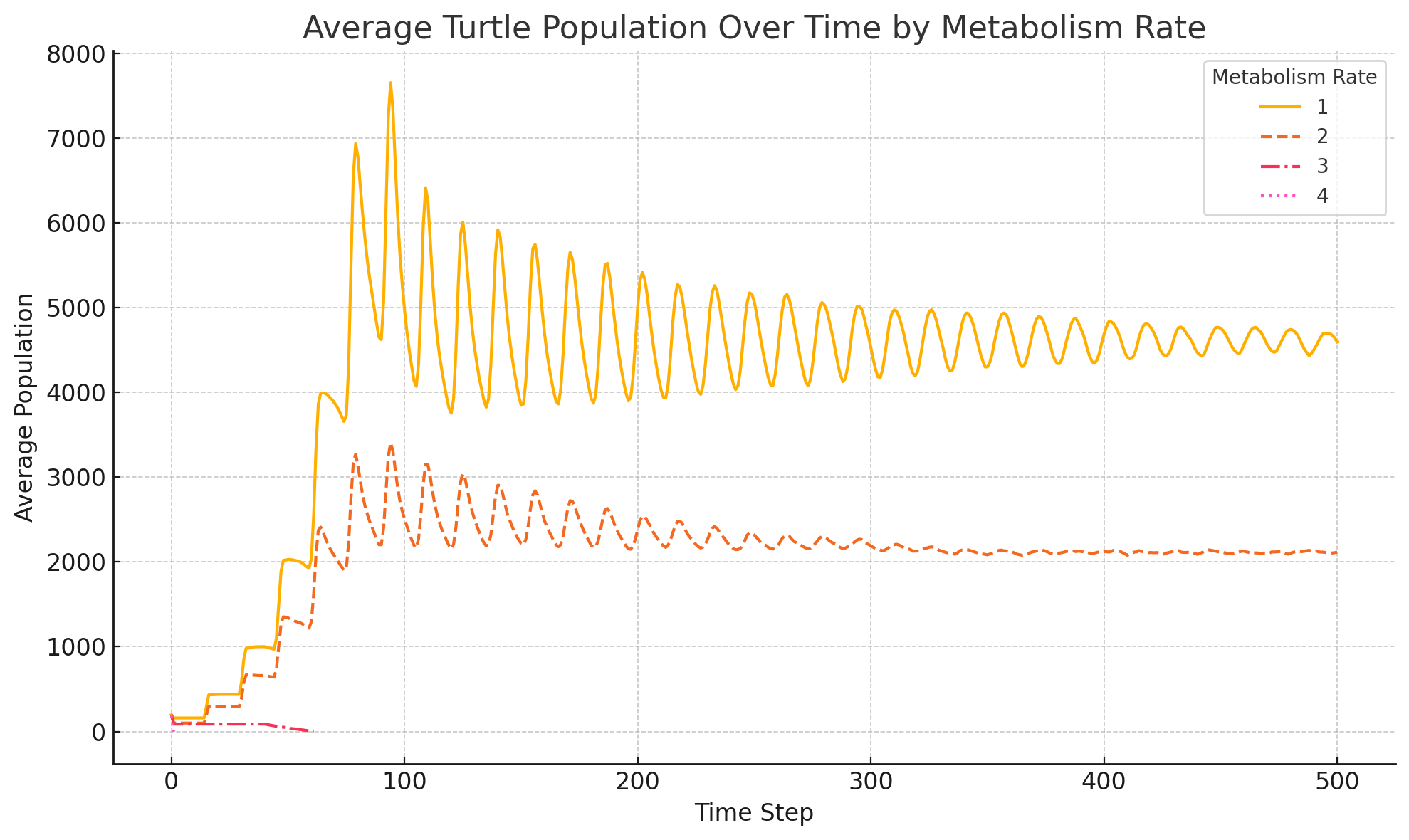}
    \caption{Metabolism‑rate }
  \end{subfigure}
  \caption{Sensitivity of population dynamics to key parameters.
           Each curve is the mean of 100 runs.
           }
  \label{fig:sensitivity}
\end{figure}

To test whether our conclusions depend on hand‑picked parameter values,
we varied one parameter at a time across the widest range that remains
archaeologically or biologically plausible and reran 100 realisations for
each setting (Figure~\ref{fig:sensitivity}).

\paragraph{(a) Donation‑rate.}
Raising the fraction of resources donated in a successful PGG
($0.10\rightarrow0.50$) increases the \emph{peak} population—because larger early
donations let more games succeed and temporarily enrich local fertility—
but has almost no effect on the \emph{long‑run equilibrium}, which
settles around $2100\pm5\%$.  Thus the early boost is transient unless a
mechanism reinforces high‑level cooperation.

\paragraph{(b) Degrade‑rate.}
Slower environmental decay (0.5) sustains fertile patches for longer and
produces a higher peak (\(\sim4\,000\) agents), whereas rapid decay (1.0)
sharply limits both the peak and the amplitude of oscillations.
Nevertheless, all curves converge on the same equilibrium, showing that
the qualitative boom–bust–stabilise pattern is robust to a two‑fold
change in ecological fragility.

\paragraph{(c) Metabolism.}
Lower metabolic demand (1 unit/tick) allows agents to stockpile sugar and
pushes the peak above 7000.  Metabolism $= 2$ reproduces our baseline.
Rates of 3 or 4 deplete resources so quickly that the population crashes
below 500 agents and never recovers, showing a clear upper bound
for viable energy requirements in this landscape. It is trivial to note that metabolisms so high have no purpose since they lead to quick extinction, which was not the case for the Aksum population. 

\medskip
Across all three levers the system exhibits the \emph{same} sequence of
phases—early boom, damped oscillations, and equilibrium—confirming that
our core finding (“short‑lived cooperation requires additional
institutions to remain effective”) is qualitatively insensitive to wide
parameter swings.  

%\subsection{ADHERENCE OF RESULTS WITH ARCHEO FINDINGS} 

%\textbf{AGGIUNGI PARAGRAFO DOVE SI PARLA DEI RISULTATI PRECISI IN RIFERIMENTO AGLI Aksum FACENDO NOTARE PRECISAMENTE QUALE è LA CORRISPONDENZA CON GLI Aksum. 
%esempio: 
%RISL 1 COME VISTO NEGLI STUDI ARCHO (STUDI X Y) SI OSSERVANO COOPERATION CLUSTERS IN THE ABM FIG2 
%RIS 2 I CLUSTER SCOMPAIONO OVER TIME AS OBSERVED IN ... FIG 3
%RIS 3 RUOLO DELL'ENVIRONMENT FIG 4 ...}

\section{Discussion}

%AGGIUNGERE RIFERIMENTI PRECISI ALLA POPOLAZIONE DEGLI Aksum. 

The findings of this study support the idea that social cooperation can emerge in pre-state communities without coercive power, but suggest that such cooperation may be inherently unstable over longer periods. In the PGG-enabled scenario, cooperation initially resulted in resource-sharing and population stability; however, without a coercive authority, the cooperative behavior declined over time: that is because when population size has reached a given threshold, free riding (that is, non-cooperative behavior) becomes evolutionarily convenient, tending to invade the system, unless a top-down intervention reverses such tendency. In sociological terms, we observe that while the population size is small enough to ensure direct social control by everyone on anyone~\cite{now98} --- mathematically, the social network can be represented by a complete graph~\cite{Erd35} --- there is no need of any sort of centralized power in order to coerce the free riders (i.e., defectors) to cooperation, for instance by means of punishment~\cite{Gintis}. Instead, once the number of members of the society is large enough to allow free riders to avoid strict social control by peers, such central power is needed to maintain a high level of cooperation~\cite{Gintis}, as it comes out from our simulations presented in previous sections. Therefore, future simulation models aimed at reproducing this evolution of societies from non-hierarchical to hierarchical structure, have to consider ingredients permitting the emergence of some kind of central power when the level of cooperation falls under a given threshold.

These results and considerations align with Stanish’s theory discussed in the Introduction that while early cooperation is achievable, sustaining it in larger populations requires centralized authority, which can impose penalties or rewards to maintain order.

\section{Conclusion}

This study provides evidence supporting the hypothesis that pre-state societies may experience cooperation without coercive power in the short term. However, it suggests that, for societies to evolve into complex states, mechanisms of coercion and punishment are likely necessary. These results provide insights into the conditions under which early complex societies may have formed, showing that, while cooperation is possible, sustainable complex social structures may depend on the emergence of centralized authority.

It also drives attention to a fundamental concept of interest to archaeology as 
an anthropological discipline, namely that social development of human groups 
is not a straightforward, unidirectional shift from simpler to more complex 
forms – i.e. from “bands” to “chiefdoms” to “states” or “kingdoms”. Such 
western evolutionary model has traditionally permeated the analysis of social 
organization in Africa’s history – including the one on the emergence, 
development and decline of Aksum’s polity –, totally ignoring the wide range of 
African alternative models for complex social organization and inter-groups 
interaction patterns. 
Moreover, the study introduces a variety of salient factors that have only rarely 
been considered in the analysis of the organization and variation in complex 
societies in Africa. Among them are the following: the very concept of “complexity”, which is 
not necessarily synonymous with hierarchization or centralization; the 
significance of ritualized economy as key element in supralocal organization of 
distantly related or unrelated people; the constant dialogue between forms of 
resistance to loss of autonomy and the structures of emergent hierarchies and 
political centralization; and the variety in the forms of power (economic, 
political, societal and ritual/ideological). Ritual power is crucial in traditional 
African societies, where a ritual leader - who is not necessarily a political leader - is recognized once credited with superior ability to make rain and ensure the 
fertility of plants, animals and woman, but having no coercive power. 
Further developments of this study will make it possible to identify the key 
factors that determined the transition from one form of social organisation to 
another and, likely, the thresholds beyond which the system needed to 
reorganize itself in order to reach a new phase of equilibrium. 
Theoretical models such as the “Internal African Frontier”, the “First Comers 
Model” and the “New Comers Model”~\cite{mcintosh1999} will be tested to reconstruct the 
dynamics of social interaction in the area of Aksum and their outcomes to verify 
which of them best fit the available archaeological record during Aksum’s 
different chronological/cultural phases. The “Segmentary State Model” will be
tested for the period of greater territorial expansion and centralized social 
organization of Aksum’s polity. The Segmentary State is held together by ritual 
suzerainity, with a centralized core and tenuous unity ~\cite{Southall}. Regional chiefs 
duplicated the king but on a smaller scale and with usually minimal formal lines 
of communication between the king and the regional chiefs, who had no 
representatives in the king’s village. The king traveled around the kingdom at 
regular intervals to collect tributes gathered by the regional chiefs. this aspect 
also matches with the content of the majority of Aksumite royal inscriptions. 
The segmentary state hypothesis may be applied to many societies in Africa 
until nominally unified by European administrations; for the Ethiopian state it is 
not new as Weissleder in 1967 studied the Amhara kingdom as a segmentary 
state~\cite{Weissleder}. 

In short, next steps of this research are by a side, finding further archaeological confirmations of the dynamics suggested by our simulations and, as we have already hinted above, conceiving new theoretical models and algorithms in order to get the emergence of a hierarchical structure of societies automatically when the external and internal conditions urge such evolution.

\

\newpage
\appendix

\section*{ APPENDIX - Pseudocode for Key Dynamics}

\begin{algorithm}
\caption{Public Goods Game (PGG) Process}
\begin{algorithmic}[1]
\For{each patch representing a proto-Aksumite site (PGG center)}
    \State \textbf{Define} radius of participation (e.g., 3 patches)
    \State Identify turtles within the participation radius

    \State \textbf{Calculate threshold for success:}
    \State \hspace{1em} base\_threshold $\gets 0.1$
    \State \hspace{1em} population\_threshold $\gets$ number of turtles $\times 0.05$
    \State \hspace{1em} resource\_threshold $\gets$ sum of 5\% of each turtle's resources
    \State threshold $\gets$ base\_threshold + population\_threshold + resource\_threshold

    \State Mark patches within the radius (visual indicator)

    \State total\_donation $\gets 0$
    \State total\_donors $\gets 0$

    \For{each turtle within the radius}
        \If{donation\_probability $>$ 0.5} \Comment{Cooperator donates}
            \State donation $\gets$ 10\% of turtle's sugar
            \State Reduce turtle's sugar by donation
            \State total\_donation $\gets$ total\_donation + donation
            \State total\_donors $\gets$ total\_donors + 1
        \EndIf
    \EndFor

    \If{total\_donation $\geq$ threshold}
        \State Increment successful PGG counter
        \State Increase fertility (max\_psugar) of patches within the radius
    \EndIf
\EndFor
\end{algorithmic}
\end{algorithm}

\begin{algorithm}
\caption{Movement and Eating}
\begin{algorithmic}[1]
\For{each turtle}
    \State Identify candidate patches within vision range with no turtles
    \State Select the patch with the highest resource (psugar)
    \State Move to the closest patch with maximum psugar
    \State Update sugar level: 
    \State \hspace{1em} sugar $\gets$ sugar - metabolism + psugar of the patch
    \State Set patch's psugar to 0
\EndFor
\end{algorithmic}
\end{algorithm}

\begin{algorithm}
\caption{Reproduction Mechanism}
\begin{algorithmic}[1]
\If{turtle's sugar $\geq$ 2 $\times$ metabolism \textbf{and} reproduction\_count $<$ 2}
    \State Hatch a new turtle (offspring):
    \State \hspace{1em} Give half of the parent's sugar to offspring
    \State \hspace{1em} Inherit parent's donation\_probability with 95\% probability
    \State \hspace{1em} Set offspring's age to 0
    \State \hspace{1em} Set random max\_age (between 40 and 60)
    \State \hspace{1em} Set metabolism to 2
    \State \hspace{1em} Set reproduction\_count for offspring to 0

    \State Increment parent's reproduction\_count by 1
\EndIf
\end{algorithmic}
\end{algorithm}

\begin{algorithm}
\caption{Resource Regrowth and Degradation}
\begin{algorithmic}[1]
\For{each patch}
    \State psugar $\gets$ max\_psugar \Comment{Regrow resources}
    \If{max\_psugar $>$ initial\_max\_psugar}
        \State Reduce max\_psugar by degradation factor (e.g., 0.7)
    \Else
        \State Reset max\_psugar to initial\_max\_psugar
    \EndIf
\EndFor
\end{algorithmic}
\end{algorithm}

\begin{algorithm}
\caption{Starvation and Aging}
\begin{algorithmic}[1]
\For{each turtle}
    \State Increment age by 1
    \If{sugar $<$ 0 \textbf{or} age $>$ max\_age}
        \State Remove turtle (dies)
    \EndIf
\EndFor
\end{algorithmic}
\end{algorithm}

\begin{algorithm}
\caption{Main Execution Loop}
\begin{algorithmic}[1]
\While{turtles exist}
    \State Run Public Goods Game (PGG) on proto-Aksumite site patches

    \For{each patch}
        \State Regenerate resources (psugar)
        \State Apply degradation to fertility (max\_psugar)
    \EndFor

    \For{each turtle}
        \State Move to the best patch (based on psugar)
        \State Consume resources (eat)
        \State Increment age
        \State Check for reproduction conditions
        \State Check for starvation or aging-related death
    \EndFor

    \State Update turtle colors based on cooperation status
    \State Update patch colors based on resource levels
    \State Advance the simulation tick
\EndWhile
\end{algorithmic}
\end{algorithm}
\end{document}